\documentclass[11pt]{article}
\usepackage{booktabs} 
\usepackage{xcolor}
\usepackage{graphicx}
\usepackage{array}
\usepackage{setspace}
\usepackage{subcaption} 
\usepackage{listings}
\usepackage{caption}
\usepackage{float}
\usepackage{authblk}
\usepackage{natbib}

\renewcommand{\arraystretch}{1.5}
\usepackage{hyperref}
\definecolor{mycolor}{RGB}{0,88,204}
\hypersetup{
  colorlinks=true,
  linkcolor=mycolor,
  urlcolor=mycolor,
  citecolor=mycolor
}
\usepackage{amsmath}
\usepackage{geometry}

\usepackage{fancyhdr}

\fancypagestyle{plain}{%
  \fancyhf{}  
  \fancyhead[R]{\thepage}  
    
}

\usepackage{mathptmx}

\usepackage{listings}
\usepackage[T1]{fontenc}
\usepackage[utf8]{inputenc}
\usepackage{amssymb}

\definecolor{keywordcolor}{rgb}{0.7, 0.1, 0.1}   
\definecolor{tacticcolor}{rgb}{0.0, 0.1, 0.6}    
\definecolor{commentcolor}{rgb}{0.4, 0.4, 0.4}   
\definecolor{symbolcolor}{rgb}{0.0, 0.1, 0.6}    
\definecolor{sortcolor}{rgb}{0.1, 0.5, 0.1}      
\definecolor{attributecolor}{rgb}{0.7, 0.1, 0.1} 

\lstnewenvironment{code}[1][]%
{
   \noindent\newline
   \minipage{\linewidth} 
   \vspace{0.5\baselineskip}
   \lstset{
 	frame = lrtb, 
 	rulecolor=\color{mycolor},
 	escapeinside={/*!}{!*/},
	language=lean, 
	aboveskip = 5mm,
	belowskip = 5mm,
	xleftmargin=2mm,
	xrightmargin=2mm,
	}
	}
{\endminipage\newline}
\lstnewenvironment{codeLong}[1][]%
{
   \lstset{
 	frame = lrtb, 
 	rulecolor=\color{mycolor},
 	escapeinside={/*!}{!*/},
	language=lean, 
	aboveskip = 5mm,
	belowskip = 5mm,
	xleftmargin=2mm,
	xrightmargin=2mm,
	}
	}
{}

\title{Reasoning about the Genetic Code in Lean}
\author{Colin T. Jones\textsuperscript{1}, Tyler R. Josephson\textsuperscript{1,2}\thanks{Corresponding author: \texttt{tjo@umbc.edu}} \\
    \textsuperscript{1}Department of Chemical, Biochemical and Environmental Engineering, \\
    University of Maryland Baltimore County, 1000 Hilltop Circle, Baltimore, MD 21250 \\
    \textsuperscript{2}Department of Computer Science and Electrical Engineering, \\
    University of Maryland Baltimore County, 1000 Hilltop Circle, Baltimore, MD 21250}
\date{\today}

\begin{document}

\maketitle

\section*{Abstract}

The genetic code exhibits redundancy, meaning that multiple distinct nucleotide sequences can encode the same amino acid sequence, a property mathematically characterized as non-injectivity. While this phenomenon is well studied in molecular biology, its rigorous verification within a computational framework remains limited. In this work, we present a machine-verified model of the central dogma of molecular biology using the Lean 4 interactive theorem prover. We formalize nucleic acid bases and amino acids as inductive types and define transcription and translation as functions mapping DNA sequences to RNA and subsequently to polypeptides. Within this framework, we prove key properties of these biological processes, including the redundancy of the genetic code, injectivity of transcription, and the non-injectivity of translation. Additionally, we formalize auxiliary processes such as DNA replication and mutation, and verify properties such as length conservation under transcription and point mutations. Finally, we demonstrate how Lean 4 helps prevent semantic errors in our code, showing its effectiveness as a tool for advancing reliable, bug-free scientific computing, particularly within the life sciences.

\section{Introduction}

Formal verification is used in computer science to provide machine-checked proofs that programs are correct \cite{deMoura2021Lean4, Song2025}. The capabilities provide unique promise to complex sciences like biology, where both software and hand-written work can be prone to error \cite{Ahmad2015}. For instance, the 2011 Nature paper, "Genomic signatures to guide the use of chemotherapeutics" was retracted due to reproducibility issues \cite{Potti2006}. The researchers used software with off-by-one indexing errors, generating false positives and rendering large swaths of their data analysis incorrect \cite{Baggerly2009}. These types of errors could be proved absent using formal methods like Lean \cite{Christiansen2023}.

In this work, we explore how genomic semantics can be represented in a theorem proving system, by developing proofs of the redundancy of the standard genetic code using Lean. Lean offers a powerful environment for encoding mathematical structures, constructing formal proofs, and verifying their correctness with machine-level precision \cite{deMoura2021Lean4, Christiansen2023, Avigad2021}.
Our goal is to bridge the gap between computational biology and formal methods by developing a verified model of the codon-to-amino acid mapping and proving key properties of this system. Specifically, we define the standard genetic code as a finite mapping within Lean, formalize the definition of redundancy in this context, and construct proofs demonstrating the non-injectivity of the genetic code. Using this framework, we also formalize DNA replication, reverse transcription, and simple mutations, along with additional proofs such as length conservation and injectivity.

From a mathematical perspective, biology is the study of dynamic systems governed by constraints, probabilities, and optimization principles. Unlike physics or chemistry, which often follow predictable laws, biology includes a level of complexity and variability and is typically stated as an inexact science that can tolerate the discrepancy between theory and reality \cite{Griesemer2013}. Still, it contains some mostly universal patterns that can be described and understood using math and proven with theorems. One of these mostly universal patterns that applies to almost all forms of life is the genetic code \cite{Nirenberg1961, Hinegardner1963, Woese1964}.

The genetic code is the set of rules by which deoxyribonucleic acid (DNA) is converted into nucleotide triplets, or codons, in messenger RNA (mRNA) and then translated into amino acid sequences of proteins. The conversion from DNA to mRNA is denoted as transcription, whereas, the conversion from mRNA to a polypeptide is denoted as translation. The direction in which information is transferred from DNA to RNA to protein is known as central dogma of molecular biology. Other important pathways relating to the genetic code include DNA replication and reverse transcription. DNA replication is used by an organism to copy and paste its genetic code in the form of DNA to preserve and spread its information through growth and reproduction. Reverse transcription in some ways violates the direction of information transfer dictated by the central dogma of molecular biology; it is a process in which information from RNA is converted to DNA.

A fundamental property of the genetic code is its redundancy, where multiple codons can encode the same amino acid \cite{Crick1966, Khorana1966, Nirenberg1965}. For example, the amino acid leucine is specified by six different codons. This redundancy is not merely an artifact of combinatorics; it plays critical roles in error minimization and evolutionary adaptation \cite{Goldman1993, Crick1966, Freeland2003, Koonin2009, Sengupta2015}. Another important property of the genetic code is the faithfulness in both forward and reverse transcription (disregarding intron processing which will be discussed) as well as DNA replication, which can be represented as length conservation. Additionally, while the entire genetic code is redundant, only the process of translation itself is non-injective; the other processes are purely one-to-one or injective.

Although the biological implications of genetic code redundancy and its other properties have been extensively studied, their formal verification within a rigorous logical and computational framework remains largely unexplored. Most of the literature focuses on systems biology or molecular signaling pathways.
Ahmad, Hasan, and Siddique formalized TP53 and other tumor related pathways in the HOL4 interactive theorem prover \cite{Ahmad2015}. Another interesting look into formal verification comes from Bahrami et al., who utilized Coq to develop a model neural network, potentially providing insight into medical applications \cite{Bahrami2018}. 

\section{Methods}
We implemented this using Lean 4, version 4.17.0-rc1, Mathlib4, and Visual Studio Code version 1.98.2. The source code is available in the \href{https://github.com/ATOMSLab/LeanGenetics}{ATOMS Lab Github}.

\section{Defining the Genetic Code}
\subsection{Informal definitions}
Nucleic acids are the fundamental components of DNA and RNA and are composed of five bases: adenine, guanine, cytosine, thymine, and uracil. Of these, thymine is unique to DNA, while uracil is found only in RNA \cite{WatsonCrick1953, Matthaei1962}. In Lean, we begin by defining these bases as objects, allowing them to serve as inputs for functions. During transcription, DNA bases are converted into their corresponding RNA bases. DNA is directional; the "forward" direction is typically denoted as 5' to 3' and the reverse direction as 3' to 5'. These labels come from the molecular structure of DNA and how it is built through condensation. DNA is also double-stranded, containing a template and a coding strand \cite{WatsonCrick1953}. The way in which transcription occurs depends on which strand is given to the computer as an input. In real life, the template strand is always read from the 3' to 5' (reverse) direction by RNA polymerase, the enzyme that catalyzes the building of the mRNA. The coding strand is complementary to the template strand; thus its sequence is equal to the mRNA sequence used in translation once all the thymines are replaced with uracils due to DNA-RNA structural differences. To represent transcription, two functions must be defined for the template and coding strands, respectively. For the template strand, we can assume the sequence is inputted in the 5' to 3' (forward) direction, so first the sequence has to be reversed mimicking RNA polymerase reading in the 3' to 5' (reverse) direction. Transcription from the coding strand can be defined as the strand itself, replacing the thymines with uracils. In translation, groups of three RNA bases on the mRNA, known as codons, are translated into amino acids according to the codon map \cite{Khorana1966, Nirenberg1965}. These bases serve as the foundation for defining transcription and translation as functions: transcription maps DNA bases to RNA bases, and translation maps lists of three RNA bases to amino acids. In other words, we want to define the five bases and two main functions: DNA-to-RNA modeling transcription and RNA-to-protein/amino acid modeling translation.

\subsection{Formal definitions}
The nucleic acids found in DNA and RNA are defined by the inductive type \texttt{NucBase}. Similarly, the amino acids are defined by the inductive type \texttt{AminoAcid} by their single letter acronyms. An inductive type is a way of defining a collection of things by starting with a few simple base cases and then giving rules for how more complex cases can be built from them. For example, in mathematics and Lean, the natural numbers are defined inductively by introducing zero as a number (our base case), and if you already have a number, you can make a new one by adding one to it. Everything in the system comes from repeatedly applying these rules, so an inductive type is essentially a structure that grows from simple beginnings through well-defined steps. This property of inductive types allows us to later define other nucleic acids like inosine. However, these other nucleic acids are not needed for the scope of this paper.

For simplicity, five nucleic bases are included in \texttt{NucBase}: uracil (U), adenine (A), guanine (G), cytosine (C), and thymine (T). Uracil and thymine are exclusive to RNA and DNA respectively. This specificity can be characterized by two functions that check any \texttt{NucBase} on whether they are a DNA or RNA base. These functions are shown in the last two lines in the code box below; \texttt{isRNABase} matches a \texttt{NucBase} for uracil, adenine, guanine, or cytosine, and \texttt{isDNABase} does the same but replaces uracil with thymine. These checks help ensure that the inputs for DNA-to-RNA and RNA-to-amino acid are DNA and RNA bases, respectively.
\begin{figure}[H]
    \centering
    \includegraphics[width=1\linewidth]{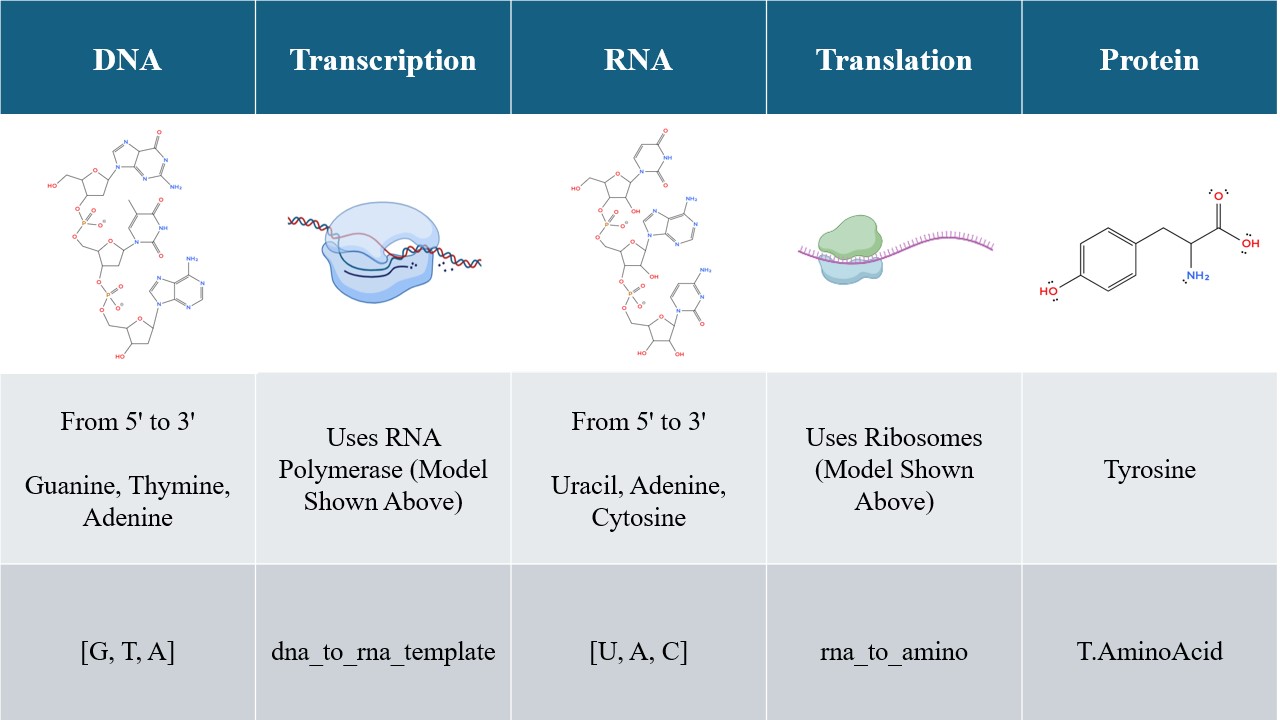}
    \caption{A demonstration of central dogma of molecular biology through the example of a trinucleotide sequence: guanine, thymine, and adenine. The central dogma of molecular biology states that information is passed in biological organisms from DNA to RNA through transcription and RNA to protein via translation. Assuming that the example given is a template strand, the RNA equivalent will be uracil, adenine, and cytosine (UAC). The UAC codon encodes tyrosine. The last row demonstrates the code equivalent in Lean. Created with BioRender.com}
    \label{fig:central-dogma}
\end{figure}
\begin{code}
/- # Definition of Nucleotide Base #
  U := Uracil
  A := Adenosine
  G := Guanosine
  C := Cytosine
  T := Thymine
-/
inductive NucBase
  | U | A | G | C | T
  deriving DecidableEq, Repr
open NucBase


/- # Definition of Amino Acid #
  Each amino acid is abbreviated by their single-letter abbreviation except for the stop codon which is represented as `STOP`.
-/
inductive AminoAcid
  | G | D | E | V | K | R | H | P | Q | F | T | W | Y | M | S | L | I | A | C | N | STOP
  deriving DecidableEq, Repr


/- # General Definitions # -/
def NucBase.isRNABase : Bool := n matches U | A | G | C

def NucBase.isDNABase : Bool := n matches T | A | G | C
\end{code}
\noindent Additionally, there is some code in between the definition of \texttt{NucBase} and \texttt{AminoAcid}. These lines are for Lean to understand how to use \texttt{NucBase}. The \texttt{open NucBase} line allows Lean to assume any instance of A, T, G, C, or U we have in our code is a part of \texttt{NucBase}. Without this line of code, we would have to explicitly assign each instance of A, T, G, C, or U as a \texttt{NucBase}. The \texttt{deriving DecidableEq, Repr} line essentially allows Lean to evaluate any statement that outputs these bases by representing them as what they currently are. The command \texttt{\#eval} is used to evaluate statements in Lean. For example, calling \texttt{\#eval U} will output \texttt{NucBase.U} in the Lean Infoview.
\begin{code}
#eval NucBase.U --Will show NucBase.U in the Infoview
\end{code}
\noindent Without \texttt{DecidableEq} and \texttt{Repr}, this call would generate an error as Lean would not know how to represent \texttt{U} as an output. Similarly, "\texttt{deriving DecidableEq, Repr}" is given to \texttt{AminoAcid} for the same reason.

Firstly, the DNA-to-RNA singlet function takes a single DNA base and converts it to its corresponding RNA base to mimic transcription. Thymine becomes adenine, adenine becomes uracil, guanine becomes cytosine, and cytosine becomes guanine.
\begin{code}
def dna_to_rna_singlet (n : NucBase) (hn : n.isDNABase) :=
  match n with
  | T => A
  | A => U
  | G => C
  | C => G
\end{code}
This function requires the input to be a DNA base through \texttt{hn}, which works as a check and stands for "hypothesis of n". In order to pass an input through \texttt{dna\_to\_rna\_singlet}, the user must prove that that input satisfies the condition of the hypothesis; for example, U would not be a valid input because it would not match with \texttt{T | A | G | C} (uracil is not a DNA base). We can use the tactics provided by Lean to prove that a particular input is a DNA base. For instance:
\begin{code}
#eval dna_to_rna_singlet A (by aesop)
\end{code}
This statement will evaluate to U and the output will be shown as \texttt{NucBase.U} in the Lean Infoview. The first part \texttt{\#eval} is a command in Lean that tells the computer to run the function. The function is called by name with an input of adenine (A). The final part, \texttt{(by aesop)}, is a proof within the \texttt{\#eval} statement. The tactic \texttt{aesop} is an advanced tactic that can automatically prove that adenine satisfies the DNA base requirement. The example above is the general structure on how to call a function for this file.

The singlet function only works for one DNA base, but DNA bases are typically arranged together. We can represent the structure of DNA by a list.
Lists work for our purpose because they are ordered sets that allow repetition of nondistinct objects.
DNA strands are ordered chemically by the 5' to 3' direction and typically contain many repeating bases. We can create a function that inputs a list of \texttt{NucBase} and outputs a list of their corresponding RNA bases, representing transcription of any strand.
\begin{code}
def dna_to_rna_template (hs : ∀ n ∈ s, n.isDNABase) : List NucBase :=
  (s.reverse).pmap dna_to_rna_singlet (by aesop)
\end{code}
The function consists of four main parts. The first line includes a different hypothesis, \texttt{hs}. This stands for "hypothesis of s" and fulfills a purpose similar to \texttt{hn}, but over the entire strand instead of one input. Breaking it down, the symbol $\forall$ stands for "for all", and $\in$ stands for "elements in". The \texttt{n} is a single \texttt{NucBase}, whereas \texttt{s} is a list of bases. In general, \texttt{hs} means: for all nucleic bases in the list \texttt{s}, they must be DNA bases. Past the hypothesis is a colon and a statement \texttt{List NucBase :=}. This statement reaffirms the type of input; in this case, \texttt{s} is a \texttt{List NucBase}. The colon and equal sign set the definition of the function. The next part of the function is its definition. First, \texttt{dna\_to\_rna\_template} reverses the DNA sequence through \texttt{s.reverse}. This mimics how RNA polymerase reads DNA in the 3' to 5' direction and assumes that the input is the template strand. Secondly, the function maps the DNA bases from the reversed list to their RNA counterparts through the function \texttt{dna\_to\_rna\_singlet} with \texttt{pmap}. The final part of \texttt{dna\_to\_rna\_template} proves that the inputs given satisfy the \texttt{hn} requirement given \texttt{hs} for the singlet function by proof: \texttt{(by aesop)}. This function assumes the template strand is given, but two similar functions can be defined for the coding strand. The difference between the coding and template strand is that the coding strand only requires replacement of the thymines with uracils to be equivalent to its complementary mRNA. It does not need to be reversed either.
\begin{code} 
def T_to_U (n : NucBase) (_ : n.isDNABase) :=
  if n = T then U else n

def dna_to_rna_coding (_ : ∀ n ∈ s, n.isDNABase) : List NucBase :=
  s.pmap T_to_U (by aesop)
\end{code}
\\
\noindent Now that an RNA sequence has been obtained, the user can convert the RNA to a sequence of amino acids---called a polypeptide---through the recursive function \texttt{rna\_to\_amino}. The recursive \texttt{rna\_to\_amino} uses a helper function \texttt{rna\_to\_amino\_single\_triplet} similar to the way \texttt{dna\_to\_rna\_template} uses \texttt{dna\_to\_rna\_singlet}. The singlet function is shown below missing some of the codons for brevity.
\begin{code}
def rna_to_amino_single_triplet (s : List NucBase) : AminoAcid :=
  match s with
  | [U, U, U] => AminoAcid.F
  | [U, U, C] => AminoAcid.F
  | [U, U, A] => AminoAcid.L
  | [U, U, G] => AminoAcid.L
  | [C, U, U] => AminoAcid.L
  ...
\end{code}
The amino acids are shown using their one-letter abbreviations; for example, in the code above, F represents phenylalanine, and L represents leucine.
\begin{code}
def rna_to_amino (L : List (List NucBase)) : List AminoAcid :=
  match L with
  | [] => []
  | y :: ys => [rna_to_amino_single_triplet y] ++ rna_to_amino ys
\end{code}
The function \texttt{rna\_to\_amino\_single\_triplet} takes a list of three RNA bases and converts it to a single amino acid. It follows the conventional genetic code; for example, \texttt{UGA} would map to a \texttt{STOP} and \texttt{GGG} would map to a \texttt{G:AminoAcid}, which represents glycine, the simplest amino acid. Combining these functions grants us the function \texttt{dna\_to\_amino\_template}, the final step of defining the central dogma of biology.
\begin{code}
def dna_to_amino_template (hs : ∀ n ∈ s, n.isDNABase) : List AminoAcid :=
  (dna_to_rna_template s hs).toChunks 3 |> rna_to_amino
\end{code}
The addition of \texttt{toChunks 3} is a coding representation of the reading frame that takes sets of three nucleotides and converts them into amino acids. The \texttt{|>} is a sequence operator in which the first output from DNA to RNA is passed to the RNA to amino acid function. This function assumes that the input given is the template strand. Another function can be defined for the coding strand.
\begin{code}
def dna_to_amino_coding (hs : ∀ n ∈ s, n.isDNABase) : List AminoAcid :=
  (dna_to_rna_coding s hs).toChunks 3 |> rna_to_amino
\end{code}

\subsection{Redundancy}
The genetic code is said to be redundant, which means that multiple distinct DNA sequences can ultimately result in the same amino acid or polypeptide during translation. In mathematical terms, this redundancy corresponds to non-injectivity, or the property of a function not being one-to-one. A function is injective if each input maps to a unique output; thus, to show that a function is not injective, it suffices to find a counterexample---two distinct inputs that yield the same output. \\

\noindent A counterexample is provided by the DNA codons AAA and GAA. If these codons are located on the coding strand of DNA, their corresponding mRNA transcripts through transcription are UUU and UUC, respectively. This follows from the rule that RNA is synthesized complementary to the template strand, but reflects the sequence of the coding strand with thymine (T) replaced by uracil (U); this operation is represented by the function \texttt{dna\_to\_rna\_coding}. \\

\noindent Next, referring to the standard codon table, both UUU and UUC are translated into the amino acid phenylalanine, represented through the function \texttt{rna\_to\_amino}. The end result is the same, despite originating from different DNA sequences. The formal proof in Lean is shown below.
\begin{figure}[H]
    \centering
    \includegraphics[width=1\linewidth]{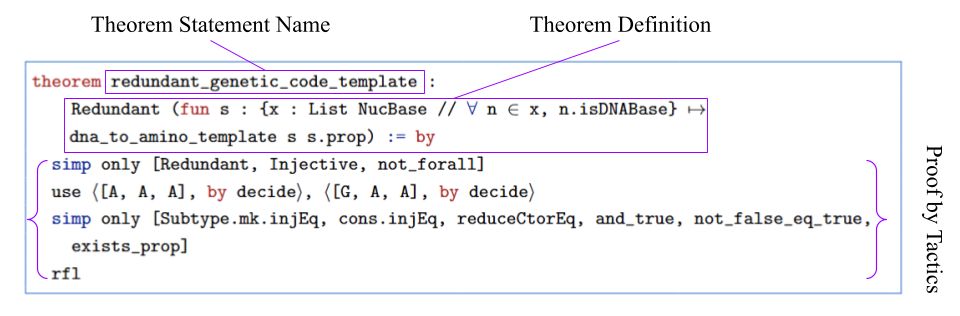}
    \caption{Lean 4 Proof of Redundancy of the Genetic Code. Sections are outlined in purple showcasing the components to a Lean proof. The Theorem Statement Name is the callback to the theorem. The Theorem Definition is the property or statement that the user is trying to prove. The body of the proof is constructed through tactics.}
    \label{fig:proof-redundancy}
\end{figure}
\noindent Proofs in Lean proceed through tactics. Tactics are commands that tell Lean how to manipulate the theorem definition. For example, \texttt{simp}, short for simplify, replaces the terms \texttt{Redundant, Injective, not\_forall} with specific logical statements. The tactic \texttt{use} then tells Lean to provide a witness, in our case, the counterexamples AAA and GAA. Then Lean calculates with \texttt{simp} and finds that transcription and translation of AAA and GAA both lead to the peptide phenylalanine by \texttt{rfl}, finishing the proof.

\noindent More proofs can be written about other functions in the genetic code. For example, length is conserved between DNA and RNA transcription. This can be proven through induction given the recursive nature of our definition of transcription. The base case for lists is an empty list represented in Lean as \texttt{[]}. An empty list of DNA bases will give an empty list of RNA bases, so their lengths are equal. Using this fact, Lean can automatically prove the following k+1 case through use of the tactic \texttt{simp}.
\begin{code}
theorem length_conservation (hs : ∀ n ∈ s, n.isDNABase = True) :
    s.length = (dna_to_rna_template s hs).length := by
  induction s
  · rfl
  · simp only [length_cons, dna_to_rna_template, reverse_cons, pmap_append, pmap_reverse, pmap_cons, pmap_nil, length_append, length_reverse, length_pmap, length_cons, length_nil, zero_add]
\end{code}
\noindent Additionally, the conversion of DNA to RNA is not redundant, instead being one-to-one or injective as every RNA base correlates to exactly one corresponding DNA base. The full proof is long and omitted in this paper, but can be found in the ATOMS Lab Github (\href{https://github.com/ATOMSLab/LeanGenetics/blob/5c34ac618471afa99e9919e5ba20da2e7289e46d/GeneticCode/GeneticCode.lean#L246C1-L246C40}{\texttt{injective\_dna\_to\_rna\_template}}). Future proofs will be mentioned in this paper, but will also be referenced this way for paper length brevity. \\

\noindent A similar proof can be written about the injectivity for the transcription of the coding strand. It is also shorter given the simpler function definition versus the template strand with no reversal and no changes in base with the exception of thymine to uracil (\href{https://github.com/ATOMSLab/LeanGenetics/blob/5c34ac618471afa99e9919e5ba20da2e7289e46d/GeneticCode/GeneticCode.lean#L278}{\texttt{injective\_dna\_to\_rna\_coding}}).

\subsection{Mutations}
Mutations are split into two main categories: point mutations and frameshift mutations. Point mutations are mutations where one DNA base is replaced with another and have three types: nonsense, where the codon now encodes a STOP; missense, where the codon now encodes a different amino acid; and silent, where the codon still encodes the same amino acid due to genetic redundancy. Frameshift mutations involve the insertion or deletion of a DNA base and change the length of the DNA strand. This causes a shift in the reading frame, hence the name.
\begin{code}
def point_mutation := s.set i n

def frameshift_delete := s.eraseIdx i

def frameshift_insert := s.insertIdx i n
\end{code}

\noindent Point mutations involve the replacement of one DNA base with another and do not change the length of the strand itself. Therefore, any DNA sequence that undergoes a point mutation conserves its length. This can be proven in Lean (\href{https://github.com/ATOMSLab/LeanGenetics/blob/5c34ac618471afa99e9919e5ba20da2e7289e46d/GeneticCode/GeneticCode.lean#L358}{\texttt{length\_conserved\_point}}).\\

\noindent Both types of mutations can be modeled using functions that act on lists. The function \texttt{set} takes an index \texttt{i} and replaces the value at that index with \texttt{n}; \texttt{eraseIdx} erases the value at \texttt{i}; and \texttt{insertIdx} inserts a value \texttt{n} at position \texttt{i}. Because they are predefined functions in Lean, they have no buggy implementations and will always work as long as the user puts in the right type. However, the definitions of \texttt{set}, \texttt{eraseIdx}, and \texttt{insertIdx} have one minor flaw that is if the index given is more than length of the function, strange results can occur. For example, \texttt{\#eval frameshift\_delete [A, G, C] 4} will return the original \texttt{[A, G, C]} list unchanged because 5 > 3. This is another example of a semantic error, an unintended issue from how we defined the code.

\subsection{DNA Replication}
DNA replication is the process by which new DNA strands are synthesized from old ones. It is essential for an organism's survival, reproduction, and propagation and is arguably the biological purpose of life. Like transcription of a template strand, it takes an old template and matches its nucleotides with their corresponding partners. The only difference is that DNA does not contain uracil, so adenine pairs with thymine. The code below is split into two parts similar to transcription. First, a helper function is defined on single nucleotides: pairing T with A and G with C. Another function then calls this function on each nucleotide in a sequence of nucleotides and maps the results into a new list.
\begin{code}
def dna_to_dna_singlet (n : NucBase) (hn : n.isDNABase) :=
  match n with
  | T => A
  | A => T
  | G => C
  | C => G

def dna_replication (hs : ∀ n ∈ s, n.isDNABase) : List NucBase :=
  s.pmap dna_to_dna_singlet hs
\end{code}
We can prove properties for DNA replication similar to those of DNA-to-RNA, including length conservation (\href{https://github.com/ATOMSLab/LeanGenetics/blob/5c34ac618471afa99e9919e5ba20da2e7289e46d/GeneticCode/GeneticCode.lean#L325}{\texttt{length\_conservation\_dna\_rep}}) and injectivity (\href{https://github.com/ATOMSLab/LeanGenetics/blob/5c34ac618471afa99e9919e5ba20da2e7289e46d/GeneticCode/GeneticCode.lean#L338}{\texttt{injective\_dna\_replication}}).
The code models DNA replication by simply translating each base on the template strand to its complement. This captures some structural properties correctly, but it abstracts away important biological details. For instance, DNA replication does not simply mirror RNA transcription with uracil swapped out for thymine; it involves complex machinery, proofreading, directionality, and synthesis of both leading and lagging strands. The prover happily accepts our definition because it is logically consistent within the formal system, but that does not mean it faithfully represents the biological process. This represents another a semantic gap whereby in the process of abstraction we have lost information of the real-life analog.

\subsection{Reverse Transcription}
A function has an inverse if and only if it is injective or one-to-one. The genetic code is redundant, not injective, so it does not have an inverse. However, the first step of the genetic code, transcription or our DNA-to-RNA function, is injective and does have an inverse. The inverse process is reverse transcription, a process especially liked by retroviruses to integrate viral DNA into host DNA \cite{Varmus1987}. We can define this in Lean and prove that it is indeed an inverse of the transcription, DNA-to-RNA function we defined before.
\begin{code}
def rna_to_dna_singlet (n : NucBase) (hn : n.isRNABase) :=
  match n with
  | U => A
  | A => T
  | G => C
  | C => G

def rna_to_dna (hs : ∀ n ∈ s, n.isRNABase) : List NucBase :=
  s.reverse.pmap rna_to_dna_singlet (by aesop)

theorem transcription_inverse_reverse_trans :
    s = rna_to_dna (dna_to_rna_template s hs)
    (by intro n₂ hn₂; have h := T_not_possible; cases n₂ <;> aesop) := by
  rw [ext_get_iff]
  constructor
  · rw [(length_conservation s hs).1]
    exact length_conserved_reverse_trans (dna_to_rna_template s hs)
  · dsimp [rna_to_dna, dna_to_rna_template]
    simp only [pmap_reverse, reverse_reverse, length_pmap, getElem_pmap]
    rintro n hn₁ hn₂
    cases n_options s[n] <;> rename_i h
    · simp_rw [h]
      have U_not_DNA : U.isDNABase = False := by
        simp only [decide_false]
        rfl
      have U_in : U ∈ s := by
        rw [← h]
        exact getElem_mem hn₂
      simp_all only [decide_false, Bool.true_eq_false]
    · obtain (h | h | h | h) := h <;> {simp_rw [h]; rfl}
\end{code}
Even with valid RNA bases (A, U, C, G), the \texttt{rna\_to\_dna} function as defined simply maps each base to its DNA complement and reverses the list. Biologically, reverse transcription is more complex: it occurs in the 5' → 3' direction, involves enzyme-driven synthesis, and may include errors such as misincorporation or template switching. The function ignores all of these realities. Lean will accept the function and proofs as logically correct---length is conserved, and the mapping is injective---but the output DNA sequence does not fully capture the biological process.

The examples above show how Lean and other theorem provers can help verify the internal consistency of code and prevent syntactic mistakes. However, they are still susceptible to semantic errors that occur when humans translate real-world concepts into formal definitions. Our definition of transcription and translation in Lean is not exempt from this either.

\begin{figure}[H]
    \centering
    \includegraphics[width=1\linewidth]{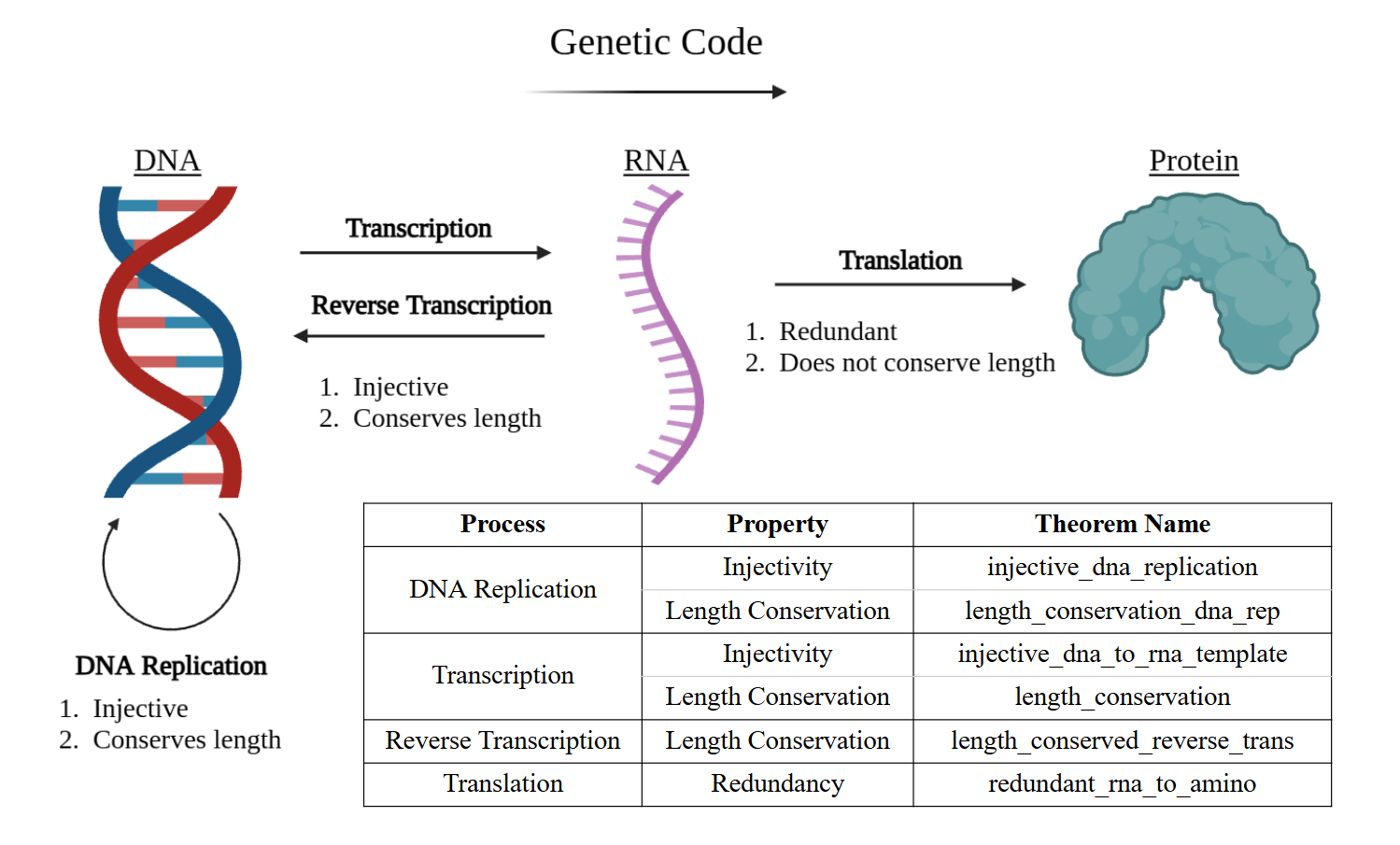}
    \caption{A diagram demonstrating the different processes in molecular biology and a table detailing each property proved and their corresponding theorem names in the code for each process.}
    \label{fig:properties-diagram}
\end{figure}
\section{Discussion}
\subsection{Challenges of Formal Biology}
Programs are susceptible to three primary types of errors: syntactic, semantic, and runtime \cite{Ugwuanyi2025}. Theorem provers mitigate both syntactic and runtime errors by enforcing rigorous computational correctness. Some specific benefits to this approach with biology are its uses in modeling mutations, replication, and evolution. Theorem provers are not invulnerable and can help protect against, but not entirely prevent semantic errors.

\begin{table}[h]
    \centering
    \renewcommand{\arraystretch}{1} 
    \setlength{\tabcolsep}{5pt} 
    \resizebox{\textwidth}{!}{ 
    \begin{tabular}{>{\raggedright\arraybackslash}p{3cm}
                    >{\raggedright\arraybackslash}p{4.5cm}
                    >{\raggedright\arraybackslash}p{5.5cm}
                    >{\raggedright\arraybackslash}p{2cm}}
        \toprule
        \textbf{Category of Error} & \textbf{Example} & \textbf{Intervention} & \textbf{Lean} \\
        \midrule
        Syntax & Not closing parentheses & Editor & Editor \\
        Runtime & Accessing element in list that doesn’t exist & Run program, program gives error message & Editor \\
        Semantic & Missing a minus sign, transposing tensor indices & Human inspection of the code; test-driven development; observing anomalous behaviour & Editor \\
        Floating-point/ Round-off & Subtracting small values from large values, ill-conditioned matrices & Modifying simulation methods, using double precision floats & --\\
        \bottomrule
    \end{tabular}
    }
    \caption{Errors in scientific computing software, and typical interventions. Our goal is to develop an approach to address syntax, runtime, and semantic errors in Lean at the ``editor'' stage, before code is compiled. Adapted from Ugwuanyi et al. (2025) \cite{Ugwuanyi2025}.}
\end{table}

\subsubsection{The STOP Codon and Accuracy}
\noindent In the real world, the STOP codon does not encode an amino acid. However, the codon table defines the STOP codon as a value of type \texttt{AminoAcid} in our context, which would be contrary to reality. Moreover, the function will not stop translation even after reaching a STOP codon as expected. The program will not necessarily start on the start codon either, defined as AUG, encoding methionine.
\begin{code}
#eval dna_to_amino_coding [A, A, A, A, T, G, T, G, A, A, A, A] (by aesop)
\end{code}
\noindent The example above evaluates to \texttt{[AminoAcid.K, AminoAcid.M, AminoAcid.STOP, AminoAcid.K]} ignoring our semantic requirement that it should start at ATG and end at UGA, UAA, or UAG, the stop codons. We can introduce a fix to our code by adding a condition for it to split the mRNA when it reaches a STOP codon, but there are still problems.
\begin{code}
def stop_dna_to_amino_coding (hs : ∀ n ∈ s, n.isDNABase) : List (List AminoAcid) :=
  ((dna_to_rna_coding s hs).toChunks 3 |> rna_to_amino).splitOnP (· == AminoAcid.STOP)

#eval stop_dna_to_amino_coding [A, T, C, T, T, T, G, G, G, T, G, A, A, A, A] (by aesop)

-- Will evaluate to [[AminoAcid.I, AminoAcid.F, AminoAcid.G], [AminoAcid.K]] split along the STOP
\end{code}
We have not taken into account RNA processing including intron and exon splicing, which changes the genetic information passed onto translation. Prokaryotes do not participate in mRNA splicing; the differences in transcription and translation between eukaryotes and prokaryotes are many. Another important difference is the presence of formyl-methionine in prokaryotes and lack of it in eukaryotes \cite{MarckerSanger1964}. The formyl group is usually cleaved off by formyl-transferase to initiate translation \cite{Dickerman1967}, but if we want to maximize accuracy, we must account for this difference and other differences between prokaryotes and eukaryotes by defining separate functions for both. To be more semantically accurate, we can also account for genetic code variations in different species, including the coding of selenocysteine and pyrrolysine. We can continue to add new conditions and extensions to this model, but biology remains an inexact science. Any model would still lack edge cases and would leave it up to the user to stop refining it. Those edge cases would only be limited to the fundamental laws of the universe. Thus, a fully formalized science of biology would essentially boil down to a fully formalized model of all of physics, something that has not yet been achieved (Hilbert's Sixth Problem \cite{Zach2023}). Being rigorous requires sacrificing some of reality to fit into a neat and well-defined framework. Here we see the difference between syntax, logic that can be verified by the computer, and semantics, something that can only be determined correct by humans or possibly now large language models.

\subsubsection{Length Conservation}

In sections 3.5 and 3.6, we proved that both DNA replication and reverse transcription conserve the length of nucleotide strands. Similarly, we did the same for forward transcription in section 3.3 while also proving the redundancy of the genetic code. However, real life biology is more complex. An organism's transcription process will differ depending on the domain to which they belong. For example, eukaryotes process their mRNA before it is exported from the nucleus. Messenger RNA undergoes three separate modifications, including: the addition of a methyl-Guanine cap to the 5' end, the splicing removal of introns, and the addition of a poly-A tail. Introns are sections of mRNA that are removed before translation through RNA splicing. On the other hand, the poly-A tail is a sequence of adenine nucleotides added to the 3' end of a mRNA to protect it from degradation outside the nucleus. Introns and the poly-A tail are especially important in the context of this paper because they modify the length of the mRNA strand so that it is incongruent with the corresponding template or coding DNA strand. The poly-A tail itself can add up to 250 adenine molecules to the end of an mRNA, but the amount that the poly-A tail adds is species specific and can vary just by chance \cite{Kuhn2009}. As far as we know, the poly-A tail does not contribute any additional coding information to the translation process; rather, it serves to protect the mRNA from degradation before it can be translated into a protein. Consequently, it does not affect the process defined here, since Lean 4 only models the transformation from DNA to protein. In contrast, the splicing of introns modifies the internal regions of the mRNA and can alter the internal length of the strand, a phenomenon that is not accounted for in the current codebase. One could implement a fix analogous to the handling of the stop codon; however, introns can vary even within the same mRNA, introducing an unexpected layer of redundancy in the transcription process (multiple finished mRNAs can be created from the same DNA strand in a process known as alternative splicing \cite{Wang2014}). This redundancy runs counter to what we have proved: the injectivity, or non-redundancy, of transcription in both the forward and reverse directions.

In both examples, biology has proven to be a particularly difficult field to formalize due to its many edge cases and conditions, which can vary even between organisms of the same species. However, efforts to formalize programs that support biological science can still yield useful results and can still prevent disastrous bugs.

\subsection{The Limits of Formalization}

Overall, we found Lean to be a sophisticated programming language with strong formal proof capabilities, but it cannot fully prevent inherent semantic errors. Nonetheless, Lean provides a meaningful layer of defense against such errors, even if they cannot eliminate them entirely and still help prevent other bugs. By including proof-writing and proof-checking in the software development process, mistakes in definitions can be identified; for example, Selsam, et al., also discuss the benefits of catching mistakes in the process writing formally-verified machine learning software \cite{selsamDevelopingBugFreeMachine2017}. 

For instance, during the process of writing the code, we incorrectly implemented the transcription of the coding strand. The coding strand is represented in the five prime to three prime direction, while RNA polymerase reads the template strand in the opposite three prime to five prime direction. The mRNA sequence matches the coding strand except that thymine is replaced by uracil. For example, if the coding strand is [T, A, G, C], the resulting RNA strand is [U, A, G, C]. However, if that was the template strand, the mRNA sequence would be [A, U, C, G].

Our first mistake was defining the transcription of the coding strand in the same way as the template strand but reversed instead of defining the coding strand as simply replacing thymine with uracil. For example, in our faulty code, a coding strand [A, G, T] would become [A, C, U] instead of correctly becoming [A, G, U]. After fixing this by correcting the definition of the coding strand, another issue was discovered when we realized that transcription of the coding strand and the template strand are only equivalent when the two strands are complementary. The lack of complementarity was fixed by including it in the theorem statement as the function \texttt{dna\_replication}, a process which is mathematically equivalent.

These were semantic errors that were unintentionally made and found only after \emph{attempting a proof} demonstrating the equivalence of transcription of a template strand and its complementary coding strand. An example is shown below of our initial incorrect definition of the template to coding equivalence and the corrected proof (\href{https://github.com/ATOMSLab/LeanGenetics/blob/ddb1752e32e8ce2230fd453e56ec3460fcad361e/GeneticCode/GeneticCode.lean#L218}{\texttt{template\_coding\_toRNA\_equivalence}}) with the revised definition.

\begin{code}
def dna_to_rna_coding (_ : ∀ n ∈ s, n.isDNABase) : List NucBase :=
  s.pmap T_to_U (by aesop)

def dna_to_rna_coding_FAULTY (hs : ∀ n ∈ s, n.isDNABase) : List NucBase :=
  s.pmap dna_to_rna_singlet hs

--The proof below does not work with the FAULTY definition due to the wrong definition in the coding strand (dna_to_rna_singlet vs T_to_U). It will also fail if the function dna_replication is removed due to loss of complementarity.
theorem template_coding_toRNA_equivalence (hs : ∀ n ∈ s, n.isDNABase) :
    dna_to_rna_template s.reverse (by aesop) = dna_to_rna_coding (dna_replication s hs) (by intro n hn; simp only [dna_replication, mem_pmap, dna_to_dna_singlet] at hn; rcases hn with ⟨n, hn, rfl⟩; aesop) := by
  unfold dna_to_rna_template dna_to_rna_coding dna_to_rna_singlet T_to_U dna_replication dna_to_dna_singlet
  simp only [reverse_reverse]
  aesop
\end{code}

Another error detected by Lean involved forgetting to reverse the direction during reverse transcription of an mRNA transcript. Like forward transcription, reverse transcription reads from three-prime to five-prime, but this was not implemented in the code. This error was revealed while attempting a proof (\href{https://github.com/ATOMSLab/LeanGenetics/blob/0b0d8265cf24dc912357e8624532cf865aa0905d/GeneticCode/GeneticCode.lean#L395}{\texttt{transcription\_inverse\_reverse\_trans}}) to show that the original DNA strand is equivalent to the reverse of the forward transcription.

Both of the examples above demonstrate the advantage of using theorem proving software such as Lean. Proofs can be constructed in Lean that are grounded in expected truths about biological processes. In the examples above, we expect that the mRNA produced from a coding strand and its complementary template strand will be the same, and that reversing transcription will recover the original DNA sequence as long as there are no introns or additional mRNA processing steps such as those that occur in pre mRNA. When these proofs cannot be constructed, they can help guide the user toward identifying issues in the code. This is because the failure of a proof often reflects an internal inconsistency within the computational framework or external inconsistency with expectations or reality.

\section{Conclusions and Outlook}
In this study, we formalized key operations of the central dogma of molecular biology using inductive types and functions in Lean. We started by defining the nucleic bases as an inductive type, \texttt{NucBase}, and then defining a condition for whether a \texttt{NucBase} qualified as a DNA base. This condition ensured that functions like \texttt{dna\_to\_rna\_template} only operate on valid inputs. We simulated transcription by reversing the input DNA (template strand) and mapping its bases to RNA complements. Translation was modeled through codon grouping through the function \texttt{List.toChunks} and conversion of each RNA triplet to its corresponding amino acid using a standard codon table. Both transcription and translation were chained to make two overarching functions --- \texttt{dna\_to\_amino\_template} and \texttt{dna\_to\_amino\_coding} --- depending on whether the user wishes to convert the template or coding strand of DNA.

We also proved biologically-relevant properties about these functions. Firstly, we demonstrated the non-injectivity or redundancy of the genetic code through the use of a counterexample. We also verified that transcription preserves the sequence length. Additional proofs explored the injectivity of DNA-to-RNA and reverse transcription and their inverse relationship, the injectivity and length conservation of DNA replication, and the equivalence between coding and template strands during transcription and translation when one strand is reversed. During this process, Lean enabled us to prevent errors in our definitions and proofs.

Along the way, we needed to make assumptions about biological scope. Nature has produced organisms with incredible diversity, including unusual specimens whose physiology does not align with the Central Dogma. Our formalization covers \emph{a model} of reality, and as such, includes assumptions which don't always hold (for instance, amino acids beyond the canonical 20, like selenocysteine). Nonetheless, \emph{all} software for biology includes assumptions like these, implicitly deciding what level of abstraction and scope they should represent. Software in any language would necessarily be more complicated to account for species-specific exceptions to the triplet $\rightarrow$ amino acid mapping (such as by duplicating functions with minor distinctions). Formal verification not only nudges the software developer to make those assumptions explicit, it also provides a toolkit for verifying when those assumptions entail certain conclusions, and when they do not.

Looking ahead, this formal framework opens the door to broader bioinformatics applications in theorem proving. Future work could extend these definitions to support RNA splicing, translation initiation and termination rules, and error-correcting mechanisms in gene expression. More broadly, any bioinformatics pipeline is a possible target for implementation in a formal environment, where formal proofs could provide correctness guarantees.

While this report and the accompanying repository were entirely human-authored (the bulk of the project being completed before general-purpose AI models could write Lean code fluently), we have observed AI becoming increasingly competent at this task. With Lean verification, AI-generated code is guaranteed to have correct syntax, but not correct semantics, a challenge we discuss, alongside the broader issues of conceptualizing formalization in science.

\section{Acknowledgments}
We thank the Lean community and Zulip for their contributions in helping build the codebase. Special acknowledgments to Matteo Cipollina for expanding on the project and offering guidance. This work was funded by NSF CAREER Award \#2236769.

\bibliographystyle{plain}
\bibliography{References}
\end{document}